\documentclass[aps,prd,twocolumn,superscriptaddress,amssymb,eqsecnum,showpacs,showkeyes,secnumarabic,graphics,floatfix,nofootinbib,tightenlines,longbibliography]{revtex4-1}
\usepackage{relsize}
\usepackage{blindtext}
\usepackage{enumitem}
\usepackage{xcolor}
\usepackage{graphicx}
\usepackage{epsfig}
\usepackage{bm}
\usepackage{cancel}
\usepackage{dcolumn}
\usepackage{amsmath}
\usepackage{hhline}
\usepackage[utf8]{inputenc}
\usepackage[breaklinks=true,colorlinks=true,
linkcolor=blue,urlcolor=blue,citecolor=cyan,
bookmarks=true,bookmarksopenlevel=2]{hyperref}

\usepackage{xcolor}

\def \beq  {\begin{equation}}
\def \eeq  {\end{equation}}
\def \ber  {\begin{eqnarray}}
\def \eer  {\end{eqnarray}}

\begin{document}
\newcommand{\newc}{\newcommand}

\newc{\be}{\begin{equation}}
\newc{\ee}{\end{equation}}
\newc{\ba}{\begin{eqnarray}}
\newc{\ea}{\end{eqnarray}}
\newc{\bea}{\begin{eqnarray*}}
\newc{\eea}{\end{eqnarray*}}
\newc{\D}{\partial}
\newc{\ie}{{\it i.e.} }
\newc{\eg}{{\it e.g.} }
\newc{\etc}{{\it etc.} }
\newc{\etal}{{\it et al.}}
\newc{\lcdm}{$\Lambda$CDM }
\newcommand{\nn}{\nonumber}
\newc{\ra}{\Rightarrow}

\title{Evading Derrick's theorem in curved space: Static metastable spherical domain wall}
\author{G. Alestas}\email{alestasg@uoi.gr}
\author{L. Perivolaropoulos}\email{leandros@uoi.gr} 
\affiliation{Department of Physics, University of Ioannina, 45110 Ioannina, Greece}

\date {\today}

\begin{abstract}
A recent analysis by one of the authors\cite{Perivolaropoulos:2018cgr} has pointed out that Derrick's theorem can be evaded in curved space. Here we extend that analysis by demonstrating the existence of a static metastable solution in a wide class of metrics that include a Schwarzschild-Rindler-AntideSitter spacetime (Grumiller metric) defined as $ds^2= f(r) dt^2 - f(r)^{-1} dr^2 - r^2 (d\theta^2 +\sin^2\theta d\phi^2)$ with $f(r)=1-\frac{2Gm}{r}+2br-\frac{\Lambda}{3} r^2$ ($\Lambda<0\; b<0$). This
metric emerges generically as a spherically symmetric vacuum solution in a class of scalar-tensor
theories\cite{Grumiller:2010bz} as well as in Weyl conformal gravity\cite{Mannheim:1988dj}. It also emerges in General Relativity (GR) in the presence of a cosmological constant and
a proper spherically symmetric perfect fluid. We demonstrate that this metric supports a static spherically symmetric metastable soliton scalar field solution that corresponds to a spherical domain wall. We
derive the static solution numerically and identify a range of parameters $m, b, \Lambda$ of the metric for which the spherical wall is metastable.  Our result is supported by both a minimization of the scalar field energy functional with proper boundary conditions and by a numerical simulation of the scalar field evolution. The metastable solution is very well approximated as $\phi(r) = Tanh\left[q (r-r_0)\right]$ where $r_0$ is the radius of the metastable wall that depends on the parameters of the metric and $q$ determines the width of the wall. We also find the gravitational effects of the thin spherical wall solution and its backreaction on the background metric that allows its formation. We show that this backreaction does not hinder the metastability of the solution even though it can change the range of parameters that correspond to metastability.

\end{abstract}
\maketitle

\section{Introduction}
\label{sec:Introduction}
Any initially static, finite energy scalar field configuration with positive definite potential energy and a canonical kinetic term in a flat 3+1 dimensional background spacetime will tend to shrink and collapse. In the presence of a negative potential energy the above scalar field configuration can remain static but it will be unstable. 

These statements are a direct consequence of Derrick's theorem \cite{Derrick:1964ww}. Derrick's theorem can be evaded by violating any of the assumptions on which it is based. For example the violation of finite energy assumption leads to the existence of global monopoles  \cite{Barriola:1989hx, Shi:2009nz, Bennett:1990xy, Harari:1990cz}, which are spherically symmetric solutions with diverging energy, whose stability is provided by their non-trivial topological properties. In a physical setup a cutoff scale is usually present and therefore global monopoles can form in physical systems. For example in a cosmological
setup  the cutoff emerges due to the cosmological 
 horizon scale while in condensed matter the cutoff
scale would be the monopole correlation scale or the size of the system. Alternatively, Derrick's theorem may be violated by introducing gauge fields in the action \cite{tHooft:1974kcl, Perivolaropoulos:1993gg, Nielsen:1973cs, Polyakov:1974ek, Perivolaropoulos:1993uj, Perivolaropoulos:2000hh} or by considering stationary \cite{Coleman:1985ki, Bazeia:2007df, Bazeia:2003qt, Perivolaropoulos:1991du, Babichev:2006cy} rather than static scalar field configurations.

The attempt to evade Derrick's theorem by violating the assumption of a flat space background has only lead to generalizations of the theorem stating that in the simplest curved spherically symmetric backgrounds (Schwarzschild and Reissner-Nordstrom) there is no static metastable finite energy scalar field configuration (soliton) \cite{Palmer1,Radmore:1978ux}. 

Therefore the following interesting questions arise: 
\begin{enumerate}
\item
Can Derrick's theorem be evaded in the presence of other non-trivial spherically symmetric background geometries leading to the existence of finite energy static scalar field configurations?
\item
If yes what are the conditions that should be satisfied by the background metric and fluid energy momentum tensor to support such configurations?
\item
What is an explicit example of a static metastable scalar field configuration that survives in a curved background but would be collapsing in a flat background?
\end{enumerate}
One of the main goals of the present analysis is to address these questions. A recent analysis \cite{Perivolaropoulos:2018cgr} by one of the authors has investigated the evolution of finite thickness topological defects \cite{Vilenkin:1984ib, Vilenkin:2000jqa, Brandenberger:1993by, Hindmarsh:1994re, Sakellariadou:2006qs} in curved space and pointed out that the violation of Derrick's theorem in curved space is possible. Here we extend that analysis by addressing all the above questions and especially questions 2 and 3. We find a metastable solution of the scalar field equations in a properly selected curved background. The solution corresponds to a static spherical domain wall in the presence of 
Scharzschild-AntideSitter metric with an additional Rindler constant acceleration term. This metric emerges generically in the vaccuum of spherically symmetric scalar-tensor theories\cite{Grumiller:2010bz}, in Weyl conformal gravity\cite{Mannheim:1988dj} and also in GR in the presence of
a spherically symmetric fluid with a black hole in its center. The gravitational effects of this scalar field configuration and its backreaction on the background metric in the context of GR are also found. 

The structure of this paper is the following: In the next section \ref{sec:Section 2} we demonstrate that Derrick's theorem can be evaded in curved space and state the condition that is required for a metastable static scalar field solution to exist in a spherically symmetric background metric. We then focus on the case of a spherical domain wall and find the necessary conditions on the background metric for the existence of metastable static spherical domain wall solution. Considering a specific metric we find the range of its parameters that satisfy these conditions. In section \ref{sec:Section 3} we minimize the energy functional and show that metastable static spherical wall solutions exist for a range of metric parameters. Using numerical simulation of dynamical field evolution we show that for proper initial conditions the spherical wall remains trapped with fixed radius in the local minimum of the energy functional. In section \ref{sec:Section 4} we discuss the gravitational effects of the derived spherical wall solution in the thin wall approximation and the backreaction on the background metric. The modification of the solution when backreaction is taken into account is also discussed. Finally in section \ref{sec:Section 5} we conclude, summarize our results and discuss possible extensions of this analysis. In what follows we use units such that the speed of light is unity ($c=1$).

\section{Evading Derrick's theorem in curved space}
\label{sec:Section 2}
Consider the spherically symmetric metric of the form 
\be
ds^2= f(r) dt^2 - f(r)^{-1} dr^2 - r^2 (d\theta^2 +\sin^2\theta d\phi^2)
\label{sphmetric}
\ee
and a canonical scalar field action of the form
\ba
S&=&\int {\cal L} \sqrt{-g}\; d^4 x \nn \\
&=&\int \left(\frac{1}{2} g^{\mu\nu}\frac{\partial \Phi}{\partial x^\mu}\frac{\partial \Phi}{\partial x^\nu}-V(\Phi)\right) \sqrt{-g}\; d^4 x
\label{sflang}
\ea
with $V(\Phi)\geq 0$. Variation of the action (\ref{sflang}) in the backround metric (\ref{sphmetric}) leads to the dynamical field equation
\be
\frac{1}{f(r)}\frac{\partial^2 \Phi}{\partial t^2}-\frac{1}{r^2}\frac{\partial}{\partial r}\left(r^2 f(r)\frac{\partial \Phi}{\partial r}\right)=-V'(\Phi)
\label{fieldeq2}
\ee
where $'$ denotes the derivative with respect to $\Phi$.
Assuming a static spherically symmetric scalar field, the components of the diagonal energy momentum tensor $T^{\mu\nu}=\partial^\mu \Phi\partial^\nu \Phi -g^{\mu \nu} {\cal L}$ are
\ba  
T_0^0&=&\frac{1}{2}f(r)(\partial_r\Phi)^2+V(\Phi)=\rho_\Phi(r)\label{density1} \\
T_r^r&=&-\frac{1}{2}f(r)(\partial_r\Phi)^2+V(\Phi)=-p_{\Phi r}(r)\\ 
T_\theta^\theta&=&T_\varphi^\varphi=\frac{1}{2}f(r)(\partial_r\Phi)^2+V(\Phi)=-p_{\Phi \theta}(r)
\ea
Thus the energy functional takes the form,
\begin{align}
\begin{split}
E &=\int d^3x \sqrt{-g}\; T_0^0 \\
&=4\pi\int_{r_1}^{r_2} \left[\frac{1}{2} f(r)\left(\frac{d\Phi}{dr}\right)^2+V(\Phi)\right]r^2dr
\label{energy1}
\end{split}
\end{align}
where the limits of integration $r_1$, $r_2$ refer to the possible existence of a black hole and a cosmological horizon respectively where $f(r)$ changes sign ($f(r)>0$ between the horizons). 

According to Derrick's theorem, the energy functional (\ref{energy1}) does not have a stable minimum in flat space when the field is rescaled by a parameter $\alpha$.  {\it Does the energy have a stable minimum in a curved space background?} In order to address this question we follow Ref \cite{Perivolaropoulos:2018cgr} and consider an initially static field configuration $\Phi(r)$ and its rescaled form $\Phi_\alpha\equiv \Phi(\alpha r)$. We search for an extremum of the energy functional (\ref{energy1}) with respect to the scaling parameter $\alpha$. Let $E_\alpha$ be the energy of the rescaled field configuration
\be
E_\alpha= 4\pi\int_{r_1}^{r_2} \left[r^2 f(r)\left(\frac{d\Phi_\alpha}{dr}\right)^2+V(\Phi_\alpha) r^2\right]dr
\label{energy2}
\ee
Setting $r'\equiv \alpha r$ and using the fact $f(r_1)=f(r_2)=0$ and the assumption $V(\Phi(r_1))=V(\Phi(r_2))=0$ (the soliton is far away from the horizons) it is straightforward to show that for the existence of a static solution a necessary condition is 
\be
\frac{1}{4\pi} \frac{dE}{d\alpha}\bigg\rvert_{\alpha=1}=I_1+I_2+I_3=0
\label{dedaeq1}
\ee
where
\ba 
I_1 &=& -\int_{r_1}^{r_2} r^3 f'(r)\left(\frac{d\Phi}{dr}\right)^2  dr   \label{i1} \\
I_2 &=& -\int_{r_1}^{r_2} r^2 f(r)\left(\frac{d\Phi}{dr}\right)^2  dr  \label{i2} \\
I_3 &=& -\int_{r_1}^{r_2} r^2 \;V(\Phi)\;  dr  \label{i3} 
\ea 
Since $I_3<0$ and $I_2<0$ we need $I_1>0$ in order to satisfy eq. (\ref{dedaeq1}) and have a static solution. Thus, the condition $f'(r)<0$ is required to hold at least for some range between the horizons. This condition can not be satisfied in a flat space where $f(r)=1$. This is consistent with Derrick's theorem. It is also not satisfied in a Schwarzschild metric ($f(r)=1-\frac{2Gm}{r}$) where $f(r)$ is a monotonically increasing function. Thus Derrick's theorem is also applicable for this metric (no static solution exists). A similar argument\cite{Radmore:1978ux,Palmer1} exists for charged Reissner–Nordström black holes where 
\be
f(r)=1-\frac{2Gm}{r}+\frac{e^2}{r^2}
\label{rnbhfr}
\ee
In this case $r_1=Gm+\sqrt{G^2 m^2+e^2}$ and $r_2=+\infty$ and as in the Schwarzschild metric $f(r)$ is monotonically increasing in the integration range leading to $I_1<0$.  Thus no static solution exists. An interesting metric where $f'(r)<0$ for some range between the horizons is the Schwarzchild-deSitter metric. Even though this metric can support static scalar field solution such solution has been shown to be unstable \cite{Perivolaropoulos:2018cgr}. Here we search for a metric with a metastable spherically symmetric finite energy scalar field solution.

Let us consider a spherical domain wall scalar field configuration of radius $r_0$ in a static spherically symmetric metric of the form (\ref{sphmetric}). The potential supporting such a configuration is the symmetry breaking potential
\be
V(\Phi)=\frac{\lambda}{4} \left(\Phi^2-\eta^2\right)^2
\label{potdomwall}
\ee
where $\eta$ is the scale of symmetry breaking. A spherical domain wall is a field configuration that interpolates between the two degenerate minima $\pm\eta$ of the potential (\ref{potdomwall}) as the surface of the wall sphere in physical space is crossed. On dimensional grounds the thickness of the domain wall is $\Delta r \simeq \lambda^{-1/2} \eta^{-1}$ and the variation of the scalar field across the wall is $\Delta \Phi=2\eta$.

In the context of the thin wall approximation the energy functional (\ref{energy1}) for a domain wall of radius $r_0$ in a background metric of the form (\ref{sphmetric}) may be easily obtained as 
\be 
\frac{E}{4\pi}\simeq r_0^2\;f(r_0)\left(\frac{\Delta \Phi}{\Delta r}\right)^2 \Delta r +  V(0)\; r_0^2 \; \Delta r
\label{energywallapprox}
\ee
which may also be written as
\be 
\frac{E(r_0)}{4\pi\lambda^{1/2} \eta} \simeq 4\;   {\bar r}_0^2\;f({\bar r}_0) + {\bar  V}(0) \; {\bar r}_0^2 
\label{energywallapprox1}
\ee 
where ${\bar r}_0\equiv \lambda^{1/2} \;\eta\; r_0$ and ${\bar  V}(0)\equiv V(0)/(\lambda\eta^4)$. In what follows we omit the bar and set $\eta\rightarrow 1$ unless otherwise specified. It is therefore a good approximation to assume that the thin wall radius evolves like a point particle in an effective potential of the form $U(r_0)=E(r_0)$ given by eq. (\ref{energywallapprox1}). Based on this approximation, we anticipate that a metastable spherical domain wall solution may exist provided the following two conditions are satisfied:
\begin{enumerate}
\item
The effective potential (\ref{energywallapprox1}) should have at least one local minimum.
\item
The metric function $f(r_0)$ should be positive at that local minimum so that it is not hidden by a horizon and no negative gradient energy (ghost) instabilities develop.
\end{enumerate}
A necessary requirement for these conditions  to be realized is that $r_0^2 f(r_0)$ should have a minimum in a region where $f(r_0)>0$ since the potential energy tension term $ V(0) \; r_0^2$ can only spoil or weaken a minimum (lower its depth). It is easy to see that the Schwarzschild metric ($r^2 f(r)=r^2-2 G m r$) satisfies the first condition but not the second and the same is true for the Schwarzschild-deSitter metric ($r^2 f(r)=r^2-2 G m r-\Lambda r^4/3$) even though in the later case there is a local maximum for $\Lambda>0$ in the potential (\ref{energywallapprox1}) \cite{Perivolaropoulos:2018cgr} indicating the presence of a static but unstable solution. 

The next simplest static spherically symmetric metric to consider is the Schwarzschild-Rindler-AntideSitter metric (also known as Grumiller metric \cite{Grumiller:2010bz,Sakalli:2017ewb,Sakalli:2014vca,Halilsoy:zva})
\be
f(r)= 1-\frac{2Gm}{r}+2\; b\; r-\frac{\Lambda}{3}r^{2}
\label{fsrads} 
\ee
which includes a linear term $2\; b\; r$ similar to the Rindler constant acceleration term\footnote{In the context of setting $\eta=1$ the constants $Gm$, $b$ and $\Lambda$ are dimensionless (we set ${\bar b}\equiv \frac{b}{\lambda^{1/2} \eta}$, ${\bar \Lambda}\equiv \frac{\Lambda}{\lambda \eta^2}$, ${\bar \Phi}=\frac{\Phi}{\eta}$ and ${\bar r}=\lambda^{1/2} \eta \; r$ and omit the bar unless otherwise specified).}. Solar system constraints have been imposed on this metric indicating that $\vert b\vert <3nm/sec^2$\cite{Carloni:2011ha,Iorio:2010tp} and it has been shown that in can lead to the production of flat rotation curves as well as contribute to the resolution\cite{Grumiller:2011gg,Iorio:2011zu} of the Pioneer anomaly \cite{Anderson:1998jd,Lammerzahl:2006ex} for $b>0$. As mentioned in the Introduction, this metric can emerge generically as a vacuum solution in  spherically symmetric scalar tensor theories \cite{Grumiller:2010bz}, and in conformal Weyl gravity\cite{Mannheim:1988dj,Mannheim:1992tr,Sultana:2012qp}. It also emerges in GR due to a spherically symmetric background fluid which could for example be attributed to nonlinear electrodynamics\cite{Halilsoy:2012sr}.
The energy momentum tensor that leads to this metric in the context of GR is diagonal with components
\ba
T_0^0&=&\rho = -\frac{4b}{\kappa r}+\frac{\Lambda}{\kappa} \label{fluidrho} \\
T_r^r&=&-p_{r} =-\frac{4b}{\kappa r}+\frac{\Lambda}{\kappa}=\rho \\
T_\theta^\theta&=&T_\phi^\phi=-p_\theta(r)=-p_\phi(r) =-\frac{2b}{\kappa r} + \frac{\Lambda}{\kappa}
\ea
where $\kappa=8\pi G$. Such fluids have been discussed in the context of relativistic stars \cite{Culetu:2011wv,Mak:2001eb}. 

It is now easy to show that 
\be
r^{2}f(r)= -2Gmr+r^2+2br^3-\frac{\Lambda}{3}r^4
\label{grm0}
\ee 
\begin{figure*}[ht]
\centering
\begin{center}
$\begin{array}{@{\hspace{-0.10in}}c@{\hspace{0.0in}}c}
\multicolumn{1}{l}{\mbox{}} &
\multicolumn{1}{l}{\mbox{}} \\ [-0.2in]
\includegraphics[scale=1.1]{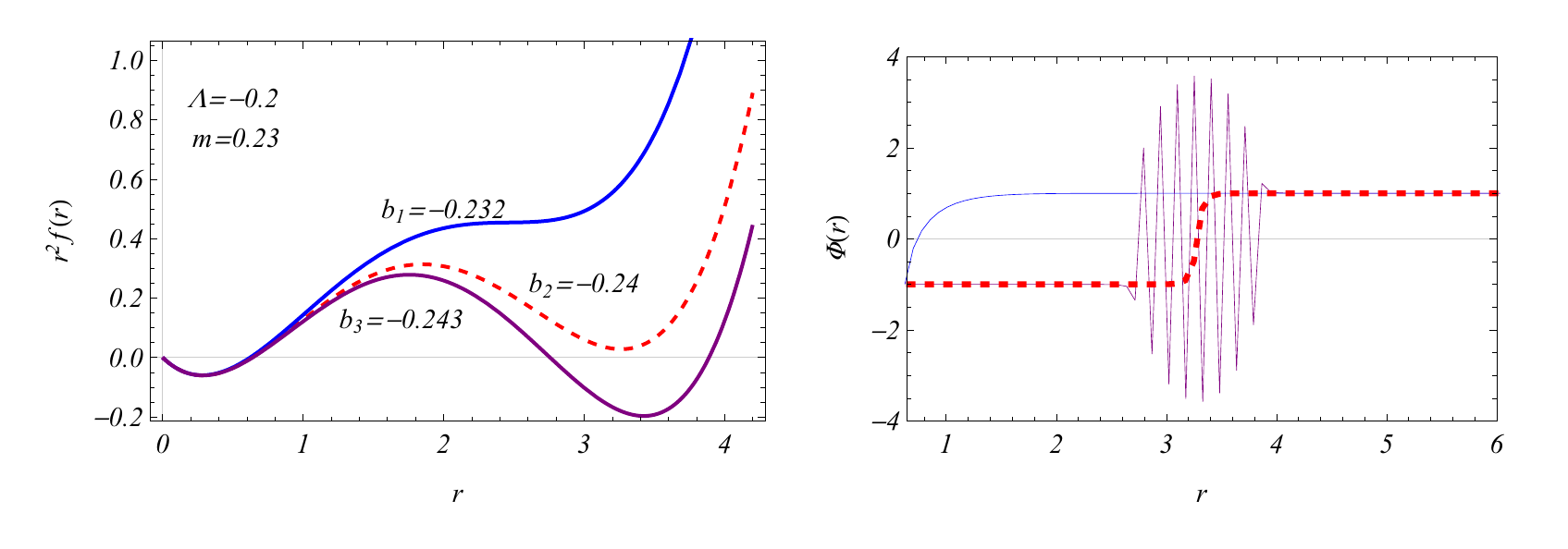} \\
\end{array}$
\end{center}
\vspace{0.0cm}
\caption{\small 
Three different behaviours of the metric function $r^2 f(r)$ and the corresponding forms of the field configuration after energy functional minimization. Only the red dotted lines correspond to metastable solutions. It is the only one where $r^2 f(r)$ has three extrema and two roots. In the other cases we get instabilities either towards collapse (blue lines) or ghost instabilities (purple lines) where the gradient terms diverge. }
\label{fig1}
\end{figure*}
\noindent can satisfy both necessary conditions stated above for the existence of a stable solutions for a range of parameters $m,b,\Lambda$. Indeed for small $r$ the linear term $-2mr$ dominates while $f(r)<0$ (inside the horizon). For intermediate $r$ the quadratic term (tension) dominates and $r^2 f(r )$ becomes an increasing function leading to the first minimum. This minimum is inside the horizon and can not lead to a metastable solution since the second condition is violated. For larger $r$ the cubic term $2br^3$ dominates and for $b<0$ it may lead to a decreasing $r^2 f(r)$ after a local maximum.  Eventually, for even larger $r$ the quartic term will dominate. For $\Lambda<0$ it will eventually lead to an increasing $r^2 f(r)$ after a local minimum at $r=r_{min}$ thus satisfying the first condition. If $b$ is not too low then this minimum will be positive thus satisfying also the second condition ($f(r_{min})>0$) and the formation of a metastable domain wall with approximate radius $r_0=r_{min}$ will be possible (see left panel of Fig. \ref{fig1}). 

The existence of such a solution can only be validated by numerical minimization of the energy functional (\ref{energy1}) for various fixed values of $m, b, \Lambda$ or by solving the static version of the field equation (\ref{fieldeq2}) and considering small perturbations around the solution. In the next section we follow the former approach.

Three representative forms of $r^2 f( r)$ are shown in Fig. \ref{fig1} (left panel). The upper curve has no local minimum ($b$ is not low enough) and thus it can not lead to a metastable solution. The middle curve has a local minimum ($b$ is low enough) and thus a metastable solution may exist if the potential tension term of the energy functional does not destroy this minimum. The lower curve has a clear minimum but the second condition is violated at this minimum  since $f( r )<0$ ($b$ is too low). Thus the solution even if it exists will suffer from gradient instabilities (ghosts) and will be hidden behind a horizon. Indeed as shown in Fig. \ref{fig1} (left panel) the  energy functional minimization leads to a metastable solution only when both conditions are satisfied (red dashed line). Otherwise, if the first condition is violated there is no local minimum and the configuration collapses (blue continues line) or if the second condition is violated, there are ghost instabilities which manifest themselves as large oscillations at the location of the minimum (purple line).

The precise range of metric parameters for which a metastable solution exists can only be found numerically by minimizing the energy functional. This range however will be a subspace of the parameter range that satisfies the two conditions necessary for metastability stated above. It is therefore interesting to identify analytically the parameter range that satisfies the two metastability conditions for the metric (\ref{fsrads}). It is easy to see that for the first condition to be satisfied $r^2 f(r)$ should have three real extrema while for the second condition, $r^2 f(r)$ (a fourth order polynomial) should have only two real roots (see the red dashed line of left panel of Fig \ref{fig1}). Thus the following two equations
\ba 
r^2 f(r)&=&-2Gmr+r^2+2br^2-\frac{\Lambda}{3}r^4=0 \label{r2fr0} \\
\frac{d(r^2f(r))}{dr}&=&-2Gm+2r+6br^2-\frac{4\Lambda}{3}r^3=0
\label{deriveq}
\ea
should have two and there roots respectively.
It is straightforward to show that for eq. (\ref{deriveq}) to have three real solutions (three extrema of $r^2 f(r)$) while eq. (\ref{r2fr0}) has only two real roots $\left| \Lambda \right|$ must be in the range
\begin{widetext}
\be
  |\Lambda |\in \left[\left|\frac{27 b\;G m-2 \sqrt{(9 b\; Gm+1)^3}+2}{18G^2 m^2}\right|,\left|-\frac{(12 b\; Gm+1)^{3/2}-18 b\; Gm-1}{18 G^2m^2}\right| \right]
\label{lrange1}
\ee
\end{widetext}
with $\Lambda<0$ and $b<0$. For $m=0$ this range becomes
\be
  |\Lambda |\in b^2  \left[3,\frac{27}{8}\right]
\label{lrange2}
\ee
The conditions (\ref{lrange1}) and (\ref{lrange2}) constitute necessary but not sufficient conditions for the existence of a spherical metastable wall solution. In the next section we show that a subspace of the above parameter range indeed corresponds to a metastable spherically symmetric wall.

\section{Minimization of the Energy Functional}
\label{sec:Section 3}

It is straightforward to show that extremization of the energy functional (\ref{energy1}), for a static
field configuration leads to the static version of the  field equation (\ref{fieldeq2}). In fact the existence of a nontrivial minimum of the energy functional implies the existence of a metastable scalar field solution. In this section we find numerically 
a range of parameters that allow for a nontrivial static scalar field configuration that minimizes the energy functional with boundary conditions that correspond to a spherical wall ($\Phi(r_1)=-1$, $\Phi(r_2)=1$).

\begin{figure}[!t]
\centering
\vspace{0cm}\rotatebox{0}{\vspace{0cm}\hspace{0cm}\resizebox{0.49\textwidth}{!}{\includegraphics{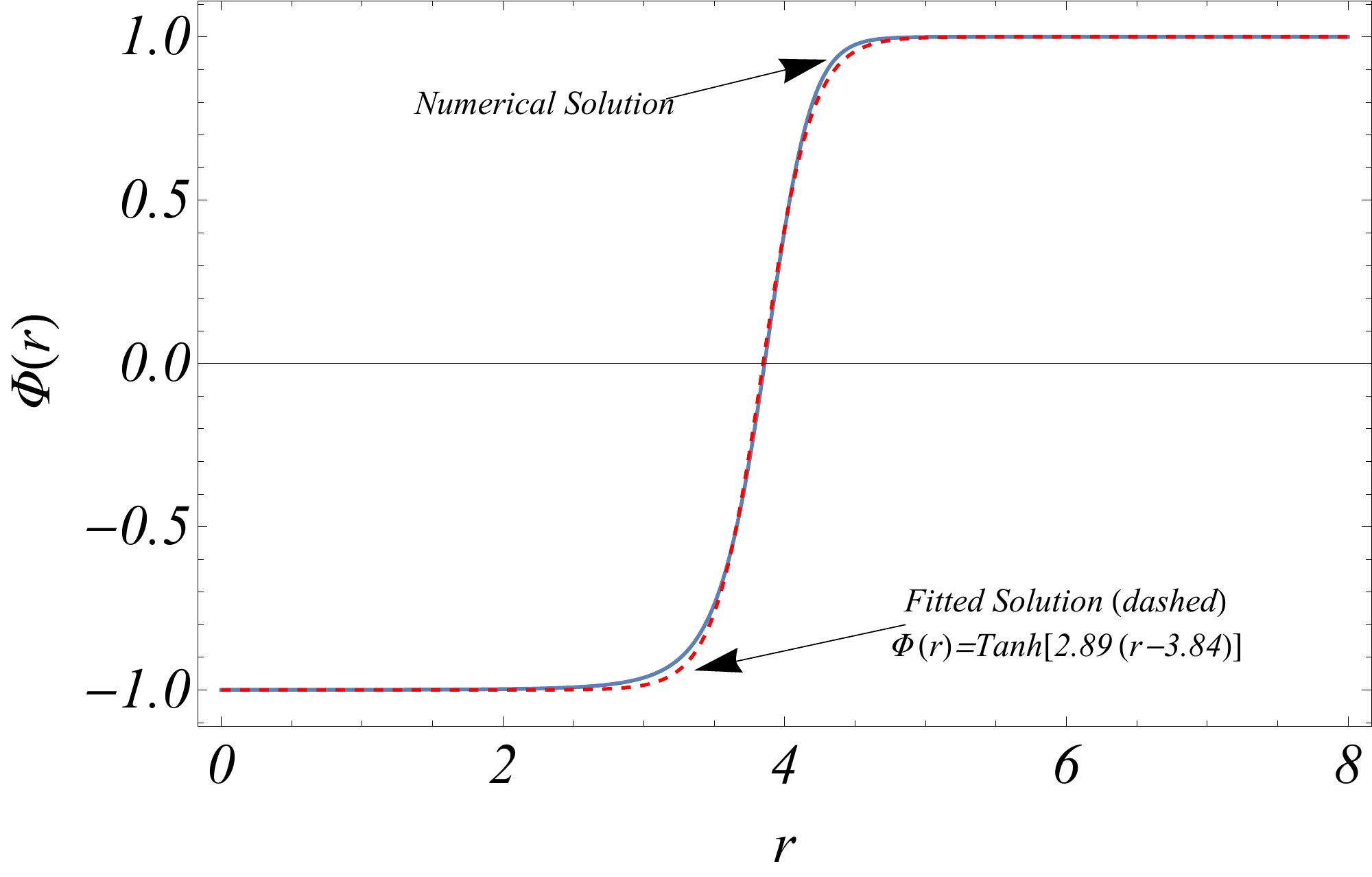}}}
\caption{The field configuration that minimizes the energy functional for $b=-0.21$ and $\Lambda=-0.14$, $m=0$ (blue line). An analytic fit (red dashed line) of the form $\Phi(r)=Tanh\left(q (r-r_0)\right)$ is also shown to provide an excellent fit to the numerically obtained metastable solution.}
\label{fig2}
\end{figure}

\begin{figure*}[ht]
\centering
\begin{center}
$\begin{array}{@{\hspace{-0.10in}}c@{\hspace{0.0in}}c}
\multicolumn{1}{l}{\mbox{}} &
\multicolumn{1}{l}{\mbox{}} \\ [-0.2in]
\includegraphics{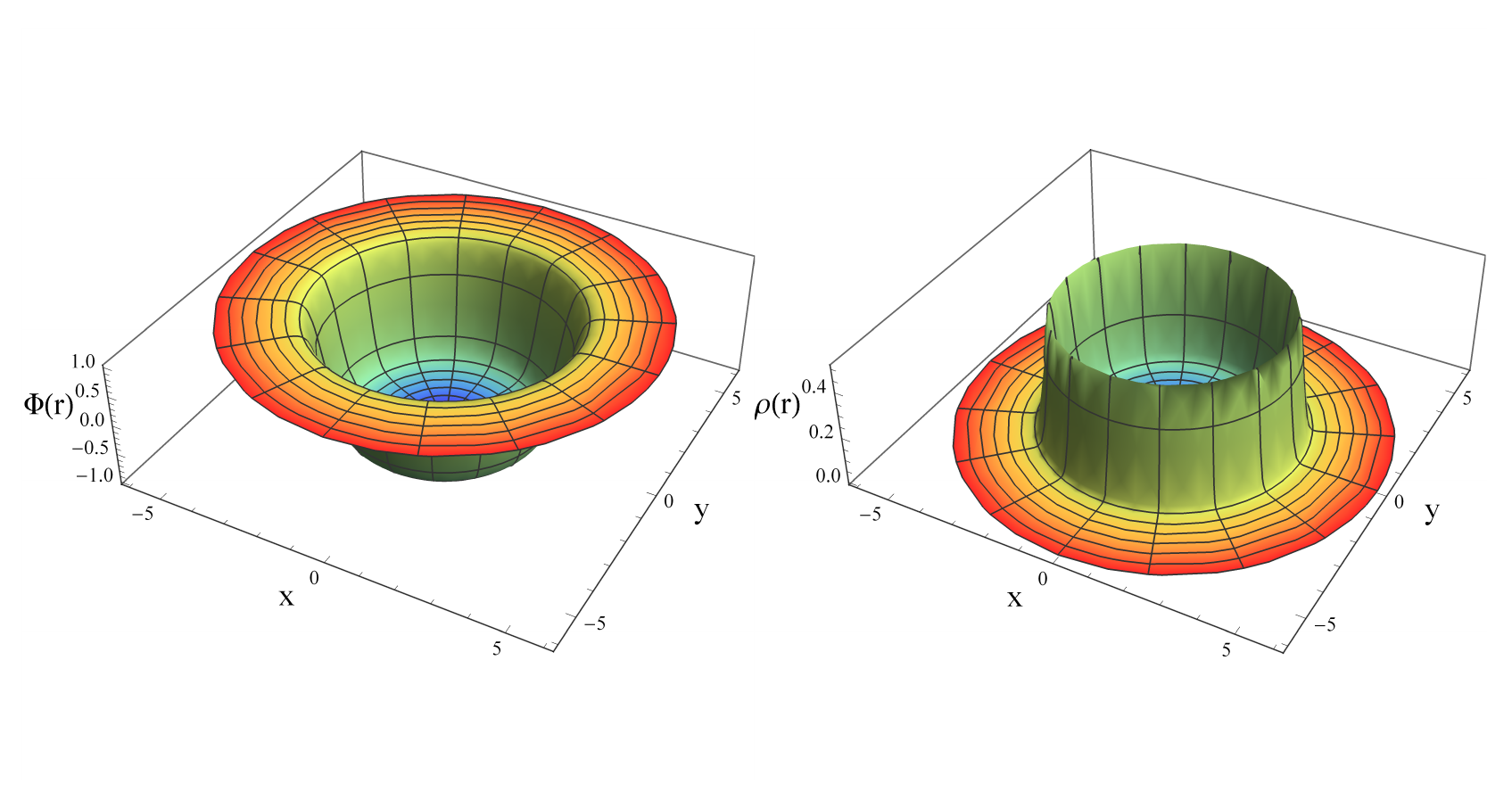} \\
\end{array}$
\end{center}
\vspace{0.0cm}
\caption{\small The field configuration of the metastable spherical wall solution for $m=0$ $b=-0.255$, $\Lambda=-0.2$ (left panel). The spherically symmetric thin shell corresponding to the energy density is also shown (right panel).}
\label{fig3}
\end{figure*}

The algorithm used to perform the energy minimization involves the following steps:
\begin{enumerate}
\item
Identify a set of parameters $m,b,\Lambda$ in the candidate range (\ref{lrange1}) with $b<0$ and $\Lambda<0$ preferably towards the lower limit of $\left| \Lambda \right|$ where the minimum of $r^2 f(r)$ is deeper. This range of parameters secures that $r^2 f(r)$ has a minimum and at the minimum we have $f(r)>0$ but it does not secure that the energy functional which includes the potential energy tension term has a nontrivial minimum. 
\item
Solve numerically the equation $r^2 f(r)=0$ to find the lowest nonzero root which is identified with the horizon $r_1$ (for $m=0$ we clearly have $r_1=0$). Note tha in the presence of only two roots there is no cosmological horizon ($r_2\rightarrow \infty$).
\item
Consider the energy functional (\ref{energy1}) with $f(r)$ given by (\ref{fsrads}) and discretize it as a sum over $N=200$ lattice points as
\ba 
&&E= 4\pi \; dr \times \nn \\ 
&&\sum_{n=0}^{N}\left[ r_i^2 f(r_i)\Phi'(r_i)^2/2+(\Phi(r_i)^2 - 1)^2/4\right]
\label{discreteenergy}
\ea
where $r_i=i \; dr + r_1$ and $dr=(r_2-r_1)/N$ and we have taken the outer boundary $r_2 \gg r_0$ where $r_0$ is the radius of the wall ($r_0$ is close to the second minimum of $r^2 f(r)$). Also we have set $\Phi(r_i)\equiv \Phi_i$ and $\Phi'(r_i)=\frac{\Phi_{i+1}-\Phi_i}{dx}$.
\item 
Minimize the sum (\ref{discreteenergy}) with respect to the field lattice values $\Phi_1,...,\Phi_{N-1}$ keeping fixed the boundary conditions $\Phi_0=-1$, $\Phi_N=+1$. This is easily done using Mathematica\cite{Mathematica,numanalysis}.
\item
Plot the interpolated field configuration that minimizes the energy functional. If the transition between the energy vaccua occurs at a radius $r_0>r_1$ then indeed a nontrivial metastable spherical wall solution exists for the considered parameter values. If the transition between the vaccua occurs at $r_0=r_1$ (the lowest $r$ boundary) then the minimum energy configuration corresponds to a collapsed configuration which could not collapse beyond $r_1$ due to the imposed boundary condition at $r_1$. Such configuration clearly does not correspond to a static solution since it implies that the energy functional has no nontrivial minimum (blue line of Fig. \ref{fig1}).    
\item
Repeat the above process scanning the parameter space to identify a range that leads to metastable spherical wall solutions.
\end{enumerate}
Using the above procedure we have found that there is indeed a finite metric parameter range for which there is a metastable spherical domain wall solution. Such a solution is shown in Fig. \ref{fig2}.  An analytic fit of the form $\Phi(r)=Tanh\left(q (r-r_0)\right)$ is also shown in the same Fig. Clearly, this ansatz provides an excellent fit to the numerically obtained metastable solution (blue continous line). The field configuration and the corresponding energy density of the metastable solution are also shown in Fig. \ref{fig3} demonstrating the spherically symmetric nature of the solution.

\begin{figure*}[!ht]
\centering
\begin{center}
$\begin{array}{@{\hspace{-0.10in}}c@{\hspace{0.0in}}c}
\multicolumn{1}{l}{\mbox{}} &
\multicolumn{1}{l}{\mbox{}} \\ [-0.2in]
\includegraphics[scale=0.55]{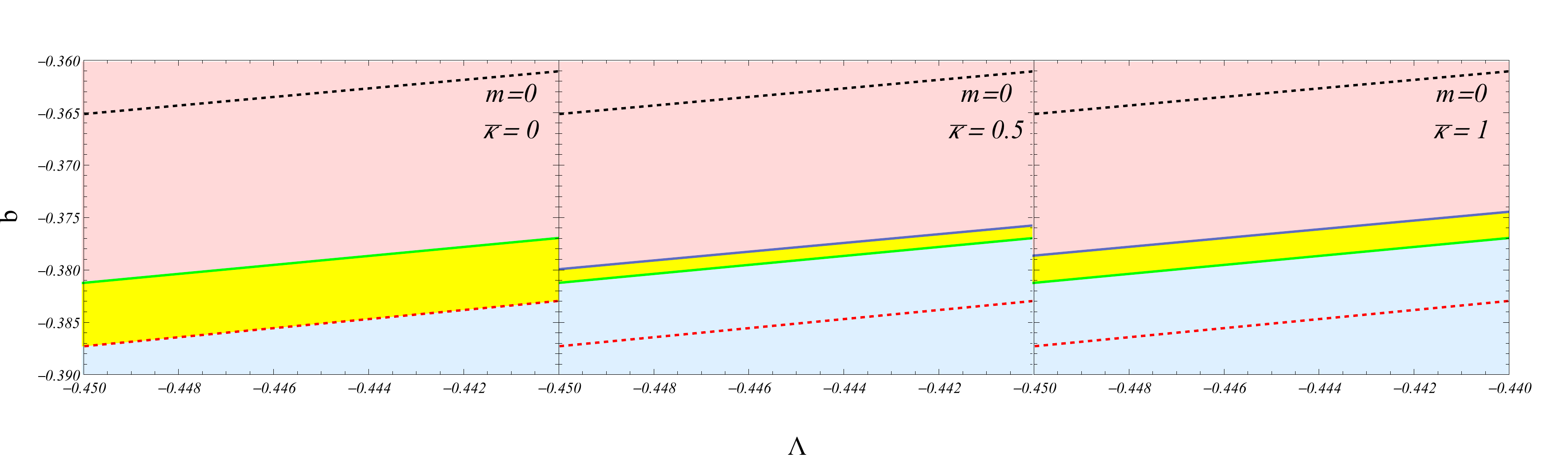} \\
\end{array}$
\end{center}
\vspace{0.0cm}
\caption{\small The yellow area corresponds to the stability region in the parameter space $b,\Lambda$ for $m=0$ where there is a minimum of the energy functional with no gradient instabilities at the location of the wall. For ${\bar \kappa}=8\pi G\eta^2 \ll \frac{\left| b \right|}{\eta}$ and ${\bar \kappa}=8\pi G\eta^2\ll \frac{\left| \Lambda \right|}{\eta^2}$ (here we restored $\eta$ for clarity)  we anticipate negligible backreaction of the wall metric on the background metric (left panel). Points above the top dashed line correspond to non existence of a minimum of $r^2 f(r)$ ($b$ is too small) while for points below the lower dashed line have a deep minimum with $f(r_{min})<0$ and thus they correspond to gradient (ghost) instabilities ($b$ is too low). The middle and left panels show how the stability region changes as the level of backreaction increases. As discussed in section \ref{sec:Section 4} backreaction tends to lower the energy minimum and lead to $f(r_{min})<0$. Thus the yellow region tends to decrease from below (see also Fig. \ref{fig9} of section \ref{sec:Section 4}). }
\label{fig5}
\end{figure*}

\begin{figure*}[!ht]
\centering
\begin{center}
$\begin{array}{@{\hspace{-0.10in}}c@{\hspace{0.0in}}c}
\multicolumn{1}{l}{\mbox{}} &
\multicolumn{1}{l}{\mbox{}} \\ [-0.2in]
\includegraphics[scale=0.55]{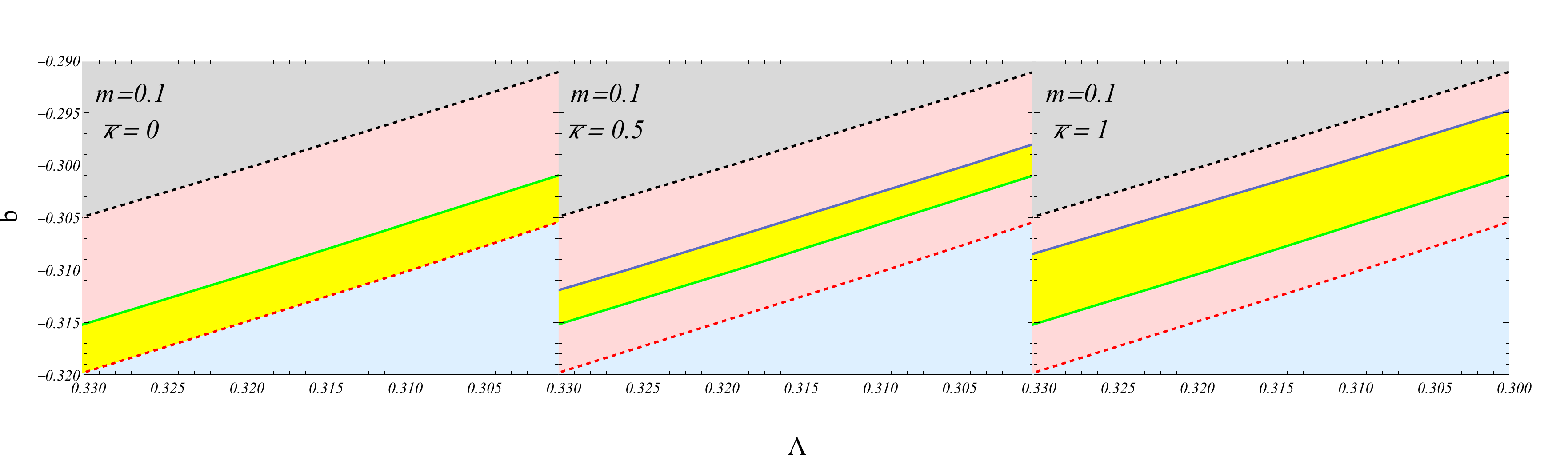} \\
\end{array}$
\end{center}
\vspace{0.0cm}
\caption{\small Same as Fig. \ref{fig5} for $Gm\eta=0.1$.}
\label{fig6}
\end{figure*}

A range of metric parameters accepting metastable spherical wall solutions is shown as the yellow region in the left panels of Figs. \ref{fig5} ($m=0$) and \ref{fig6} ($Gm\eta=0.1$). These panels are based on the assumption that backreaction of the wall metric on the background metric is negligible. The middle and right panels of these figures show the deformation of the stability region in the presence of backreaction expressed through the dimensionless parameter ${\bar \kappa} \equiv 8\pi G \eta^2$. In the next section we evaluate the effects of backreaction which become important when ${\bar \kappa}$ becomes comparable with the dimensionless parameters  $\frac{\left| b \right|}{\eta}$ and $\frac{\left| \Lambda \right|}{\eta^2}$.

\begin{figure*}[ht]
\centering
\begin{center}
$\begin{array}{@{\hspace{-0.10in}}c@{\hspace{0.0in}}c}
\multicolumn{1}{l}{\mbox{}} &
\multicolumn{1}{l}{\mbox{}} \\ [-0.2in]
\includegraphics[scale=0.55]{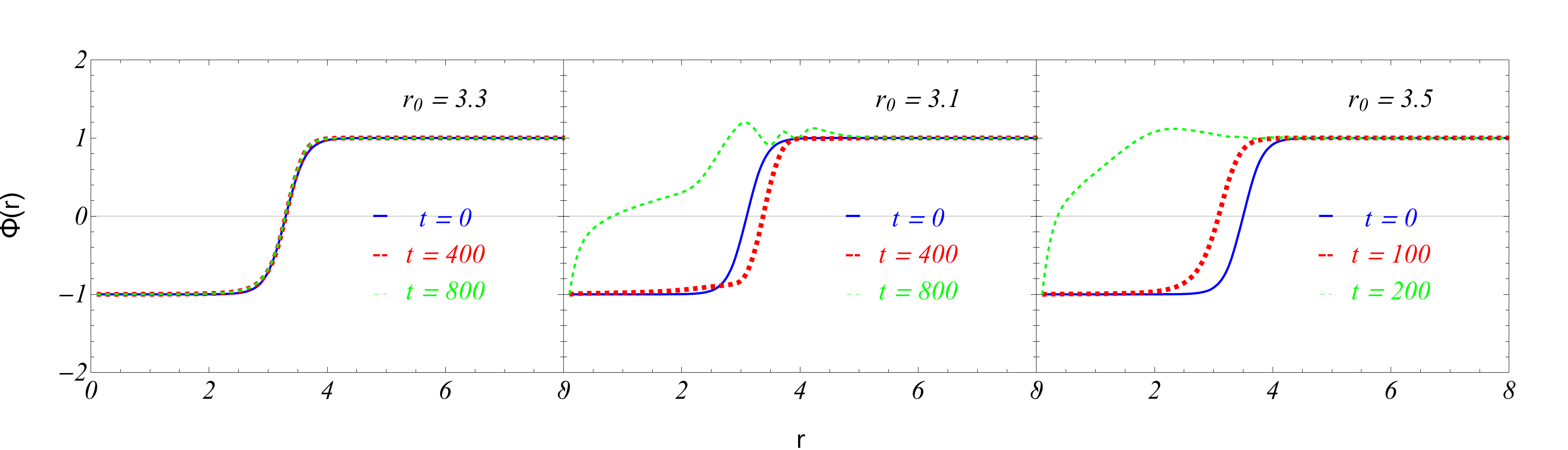} \\
\end{array}$
\end{center}
\vspace{0.0cm}
\caption{\small Simulation of the field evolution with parameter values $m=0$, $b=-0.25$, $\Lambda=-0.2$ and initial wall approximated by $Tanh\left[q(r-r_0)\right]$ with initial radius $r_0$ equal to the radius to the static solution (left panel $r_0=3.3$), slightly smaller (middle panel $r_0=3.1$) and slightly larger (right panel $r_0=3.5$) than the radius of the static solution. The initial configuration remains static in the left panel but it collapses in both other panels.}
\label{fig7}
\end{figure*}

The existence of the spherical wall metastable solution may also be demonstrated by numerical simulation of field evolution obtained by solving numerically the dynamical field equation (\ref{fieldeq2}) with initial conditions close to the metastable solution obtained by minimization of the energy functional. Thus we can obtain the range of the initial conditions that get trapped at the metastable solution and visualise the general form of the evolution of the spherical wall. The time evolving scalar field configuration for three characteristic times and three different initial conditions is shown in Fig. \ref{fig7}. In the left panel we show the time evolution of the static spherical metastable wall solution obtained by minimizing the energy functional (\ref{energy1})  with parameter values $m=0$, $b=-0.25$, $\Lambda=-0.2$. For the initial condition, we have used a fit of this solution by the $Tanh\left[q \left(r-r_0\right)\right]$ function which provides an excellent fit (see eg Fig. \ref{fig2}). As expected the evolution leads to no change of the initial configuration for arbitrarily long time of evolution. In the middle panel we show the evolution of a spherical wall slightly shifted to the left with respect to the static solution.The wall initially slightly expands moving outwards to the right towards the energy minimum (red dashed line) but eventually it shrinks and collapses (green dotted line) as it can not get trapped at the energy minimum. Similarly when the initial wall has a radius larger than the static solution it initially shrinks towards the radius of the static solution (energy minimum) where it delays its evolution until it eventually collapses (green dotted line at the right panel). 

The above discussion and existence of an energy functional minimum and static metastable solution can be generalized for any metric of the form (\ref{sphmetric}) with a power series $f(r)$.
\be 
f(r)=1-\sum^{N}_{n=-N}a_{n}r^{n}
\label{fpoly}
\ee
The Einstein tensor corresponding to this metric is,
\begin{widetext}
\be
G^{\mu}_{\nu}=\sum^{N}_{n=-N}
  \begin{bmatrix}
    a_{n}(n+1)r^{n-2} & 0 & 0 & 0 \\
    0 & a_{n}(n+1)r^{n-2} & 0 & 0 \\
    0 & 0 & \frac{1}{2}a_{n}n(n+1)r^{n-2} & 0 \\
    0 & 0 & 0 & \frac{1}{2}a_{n}n(n+1)r^{n-2}
  \end{bmatrix}
\ee
\end{widetext}
Therefore, the energy - momentum tensor supporting the metric function (\ref{fpoly}) is 
\begin{align}
T^{0}_{0} &=\frac{1}{\kappa}\sum^{N}_{n=-N}a_{n}(1+n)r^{n-2}=\rho \label{rho}\\
T^{r}_{r} &=T^{0}_{0}=-p_{r} \label{p_r}\\
T^{\theta}_{\theta} &=\frac{1}{2\kappa}\sum^{N}_{n=-N}a_{n}n(1+n)r^{n-2}=-p_{\theta} \label{ptheta}\\
T^{\phi}_{\phi} &=T^{\theta}_{\theta}=-p_{\phi} \label{pPhi} 
\end{align}
 As expected the term $n=-1$ corresponds to zero energy momentum term (vacuum solution) while for $n=2$ we obtain the cosmological constant term (constant energy density-pressure) and for $n=0$ we have the case of a global monopole (zero angular pressure components and energy density, radial pressure $\sim r^{-2}$). We could now consider polynomial forms of $r^2 f(r)$ and for each local minimum with $f(r_{min})>0$ we could identify new static metastable wall solutions following the same method as for the Grumiller metric (\ref{fsrads}). The systematic study of this general class of solutions is an interesting extension of our analysis.

\section{Gravitational effects of Wall solution - Backreaction on metric}
\label{sec:Section 4}

A crucial assumption made in the above derivation of the static solution is that the backreaction of the wall energy density on the background metric is negligible. Here we quantify the implications of this assumption on the parameters of the background metric and the scalar field. For backreaction to be negligible, the energy density of the scalar field should be much smaller than the energy density of the background fluid. Thus we demand
\be 
\rho_\Phi \ll \rho_b + \rho_\Lambda = \frac{-4b}{\kappa r}+\frac{\Lambda}{\kappa}
\label{bcrcond}
\ee
where $\rho_\Phi$ is the scalar field energy density given by eq.(\ref{density1}). For the potential (\ref{potdomwall}) the total energy density takes the form
\begin{widetext}
\be
\rho_{tot}=\frac{\lambda \eta^4}{\bar \kappa}\left[\bar{\kappa}\left[\frac{1}{2}f(\bar{r})(\partial_{\bar{r}}\bar{\Phi})^{2}+\frac{1}{4}(\bar{\Phi}^{2}-1)^{2}\right]-\frac{4\bar{b}}{\bar{r}}+\bar{\Lambda}\right]
\label{totdens}
\ee
\end{widetext}
where $\bar \kappa=8\pi G \eta^2$, ${\bar b}\equiv \frac{b}{\lambda^{1/2} \eta}$, ${\bar \Lambda}\equiv \frac{\Lambda}{\lambda \eta^2}$, ${\bar \Phi}=\frac{\Phi}{\eta}$ and ${\bar r}=\lambda^{1/2} \eta \; r$. Thus for ${\bar r}>1$ the requirement for negligible backreaction indicates that 
\ba 
{\bar \kappa}&\ll& {\bar b} \label{bcrcond1} \\
{\bar \kappa}&\ll&{\bar \Lambda}
\label{bcrcond2}
\ea
It is straightforward to re-derive the static metastable domain wall solution taking also into account the effects of backreaction of the wall energy momentum tensor on the background metric. This task involves the following steps:
\begin{enumerate}
\item
Assume a metric of the form (\ref{sphmetric}) and set
\be 
f(r)=1-g(r)
\label{fgrel}
\ee
The Einstein equation for the energy density $G_0^0=\kappa T_0^0=\kappa \rho_{tot}$ takes the form
\begin{widetext}
\be
\frac{g^\prime(\bar{r})}{\bar{r}}+\frac{g(\bar{r})}{\bar{r}^{2}}=\bar{\kappa}[\frac{1}{2}f(\bar{r})(\partial_{\bar{r}}\bar{\Phi})^{2}+\frac{1}{4}(\bar{\Phi}^{2}-1)^{2}]-\frac{4\bar{b}}{\bar{r}}+\bar{\Lambda}
\label{ee00}
\ee
\end{widetext} 
For $\bar \kappa=0$ (no backreaction) the solution of (\ref{ee00}) leads to the background metric function (\ref{fsrads}). 
\item
Using the unperturbed background metric (\ref{fsrads}) for a set of metric parameters we minimize the energy functional and find the static metastable wall solution. We then use it to evaluate the scalar field energy density $\rho_\Phi (r)$ which is the factor multiplying $\bar \kappa$ in eq. (\ref{ee00}).
\item
Fix $\bar \kappa$ and use the evaluated scalar field energy density to solve the Einstein equation (\ref{ee00}) with boundary condition $g(r_1)=1$ ($r_1$ is the radius of the inner horizon where $f(r_1)=0$) to find $g(r)$ which includes the effects of the wall energy density.
\item
Use the derived metric function search for a minimum the energy functional (\ref{discreteenergy}) and if it exists derive the new static metastable solution which now includes the effects of backreaction. In the limiting case of an infinitely thin domain wall solution with energy density
\be
\rho_\Phi = \frac{E_\Phi}{4\pi r_0^2} \delta(r-r_0)
\label{rhothinwall}
\ee
(where $E_\Phi$ is the total energy of the wall) eq. (\ref{discreteenergy}) is easily solved and leads to the metric function
\be
f(r)=1-\frac{2m}{r}-\frac{2E_\Phi \Theta(r-r_0)}{r}+2br-\frac{\Lambda}{3}r^2
\label{metricthinwall}
\ee
which is consistent with Birkhoff's theorem.
\item
Repeat the above steps with different metric parameters at step 2, to identify the metric parameter region for which a metastable solution exists including the effects of backreaction. The new parameter region of stability is shown in the middle ($\kappa=0.5$) and right ($\kappa=1$) panels of Figs. \ref{fig5} ($m=0$) and \ref{fig6} ($Gm\eta=0.1$).
\end{enumerate}

\begin{figure}[!t]
\centering
\vspace{0cm}\rotatebox{0}{\vspace{0cm}\hspace{0cm}\resizebox{0.49\textwidth}{!}{\includegraphics{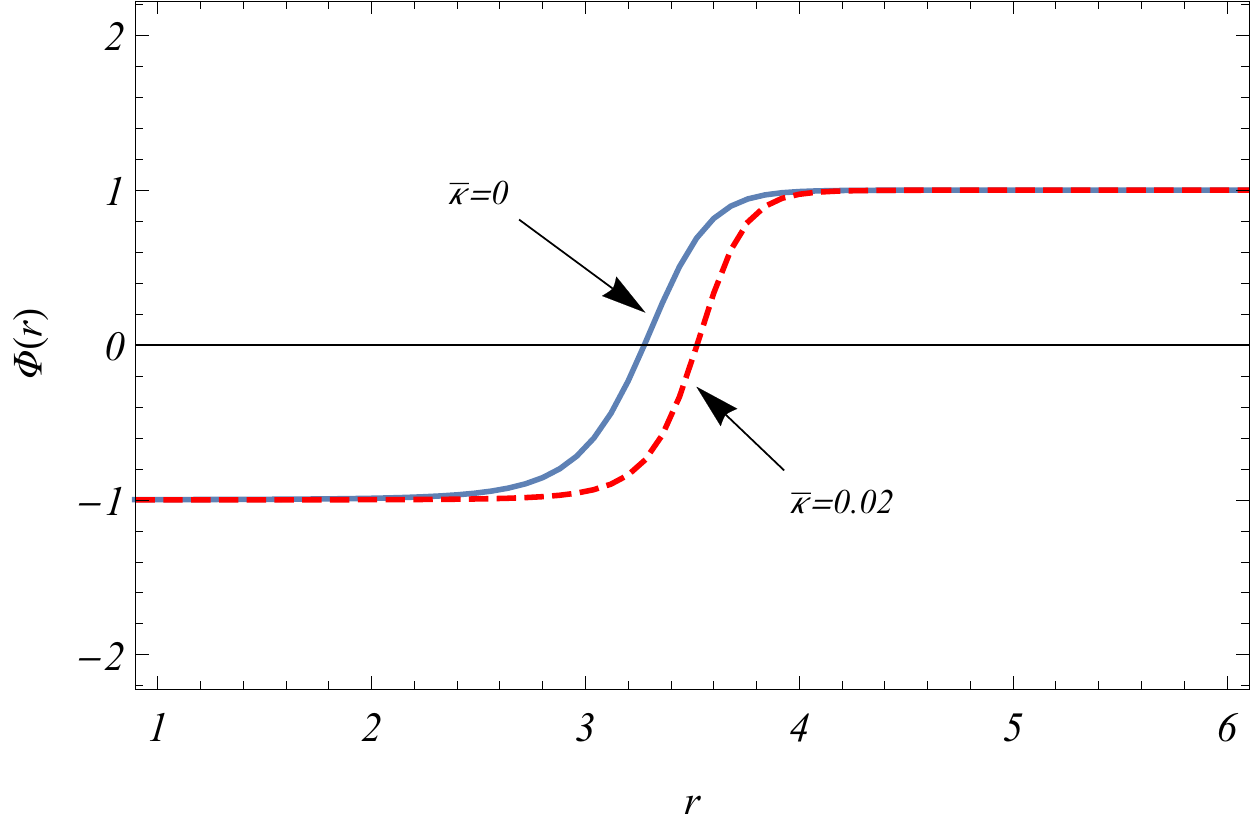}}}
\caption{The static wall solution in the absence and in the presence of backreaction. Notice that backreaction changes not only the depth but also the position of the minimum of the energy functional (see also \ref{fig9}).}
\label{fig8}
\end{figure}
\begin{figure}[!t]
\centering
\vspace{0cm}\rotatebox{0}{\vspace{0cm}\hspace{0cm}\resizebox{0.49\textwidth}{!}{\includegraphics{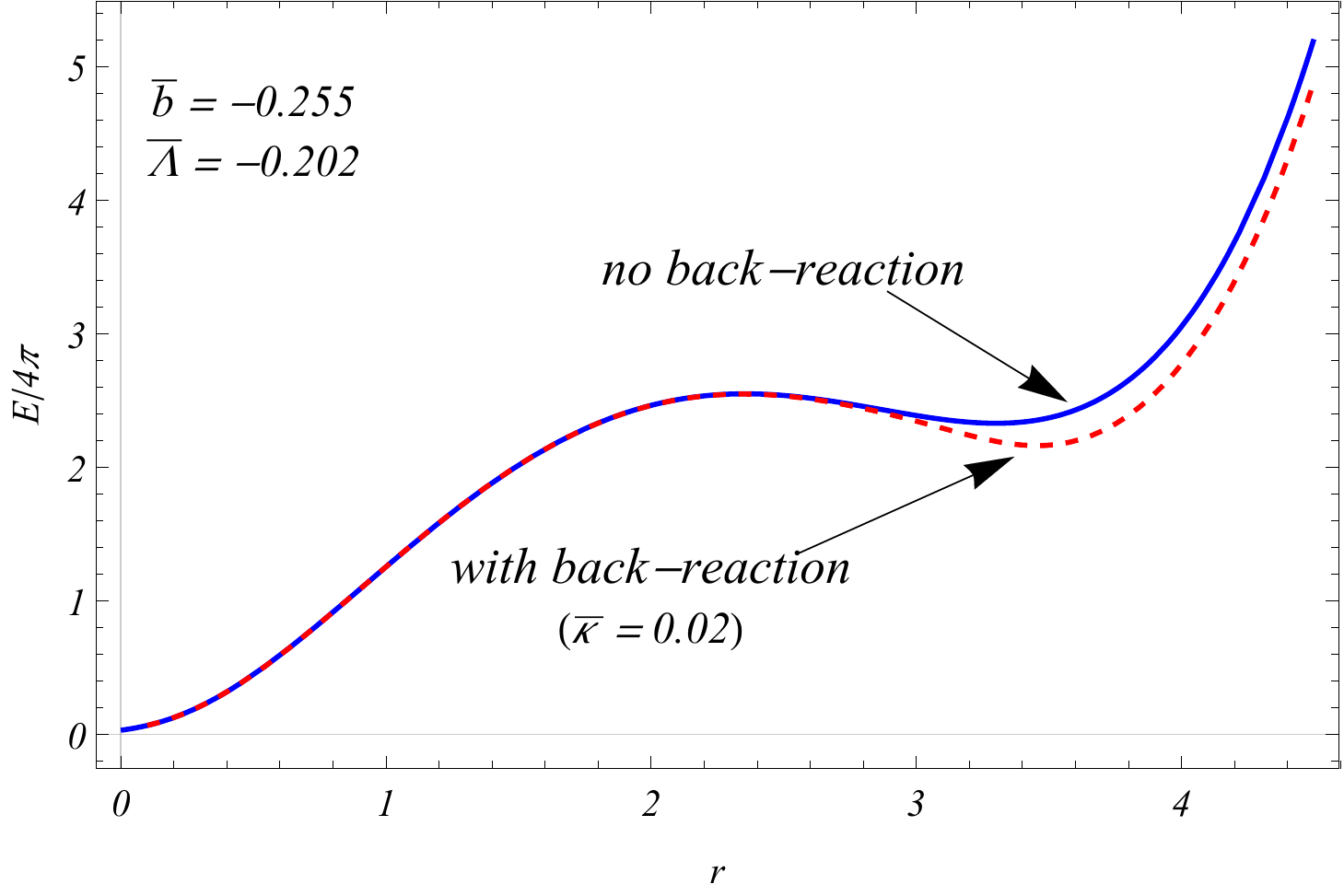}}}
\caption{The effects of a small backreaction (${\bar\kappa}=0.02$ on the static solution and on the energy functional. We have used the parameter values $m=0$,  $b = -0.25$, $\Lambda = -0.2$.}
\label{fig9}
\end{figure}

The effects of a small backreaction (${\bar\kappa}=0.02$ on the static solution and on the energy functional are shown in Figs. \ref{fig8} and \ref{fig9} respectively. In evaluating the energy functional we approximated the field configuration as $\Phi(r)=Tanh\left[3\left(r-r_0\right)\right]$

\section{Discussion-Outlook}
\label{sec:Section 5}
We have shown that Derrick's theorem can be evaded in curved space leading to finite energy static metastable scalar field configurations. We have also found an explicit example where this violation occurs leading to the formation of a static metastable spherical domain wall in a Schwarzschild-Rindler-AntideSitter background space. We have shown that backreaction effects do not destabilize the solutions even though they change the range of metric parameters where the wall is metastable. By generalizing this background metric, an infinite number of such solutions may be found for all metrics that have a metric function $f(r)$ such that $r^2 f(r)$ has a minimum at a point where $f(r)>0$.

As mentioned above, the background metric (\ref{sphmetric})-(\ref{fsrads}) considered in our analysis may emerge in the context of GR by a background fluid with energy density given by (\ref{fluidrho}). It may also emerege in a more generic manner as a vacuum solution in IR gravity in the context of an effective spherically symmetric scalar tensor theory \cite{Grumiller:2010bz}, or as a vacuum solution of Weyl conformal gravity\cite{Mannheim:1988dj}. It emerges naturally as a vacuum solution in  spherically symmetric scalar-tensor theories whose $t-r$ subspace is described by the action
\be
S=-\frac{1}{\kappa}\int d^4 x \sqrt{-g}\left[\Phi^2\; R +(\partial \Phi)^2-  V(\Phi)\right]
\label{graction}
\ee
with
\be
V(\Phi)= \Lambda \Phi^2+b\Phi+c+O(1/\Phi)
\label{vphigr}
\ee
where $\Lambda$, $b$ and $c$ are constants ($c$ can be set to 1 by a rescaling of $\Phi$). The potential in the context of this spherically symmetric scalar tensor theory is constrained to have terms up to quadratic order in $\Phi$ since higher order terms would lead to a curvature singularity for large $\Phi$\cite{Grumiller:2010bz}. Also the terms $O(1/\Phi)$ produce subleading contributions in the IR limit (large $r$). Finally the form of the nonminimal coupling $\Phi^2 R$ is also generic as any other choice of the power of $\Phi$ would not reproduce the Newtonian potential $\sim -M/r$ at smaller $r$. Thus, this is a generic action in this class of theories with a corresponding generic vacuum solution (\ref{fsrads}). This vacuum metric, for $b>0$ has been shown to reproduce well the flat galactic rotation curves \cite{Lin:2012zh} as it leads to an effective potential for the motion of massive particles of the form\cite{Halilsoy:zva}
\be
V_{eff}=-\frac{GM}{r}+\frac{l^2}{2r^2}-\frac{GMl^2}{r^3}-\frac{\Lambda}{6}r^2+br\left(1+\frac{l^2}{r^2}\right)
\label{veff}
\ee
where $l$ is the angular momentum of the massive particle. The term $b\;r$ gives rise to the constant Rindler acceleration which if positive (attractive) can play the role of dark matter in the galactic rotation curves.

If the derived metastable spherical wall forms on cosmological scales it may produce interesting cosmological observational signatures including a characteristic lensing pattern\cite{Sultana:2012zz, Li:2011ur, Faber:2005xc} as well as possible glitches in the galactic rotation curves. The investigation of such observational signatures is an interesting extension of this project.
The derivation of similar solutions in systems with axial or planar symmetry is also an interesting extension of this analysis.

\textbf{Numerical Analysis Files:} The Mathematica files used for the numerical analysis of this study and the construction of the figures are publicly available\cite{numanalysis}.


\bibliography{bibliography}

\end{document}